\documentstyle[12pt,epsf]{article}
\newcommand{\al}{\alpha}
\newcommand{\be}{\beta}
\newcommand{\ep}{\epsilon_{\mu \nu \alpha \beta}}

\begin{document}

\begin{titlepage}
\begin{flushleft}
MZ-TH/94-25\\
December 1 1994
\end{flushleft}
\begin{center}
\large
\bf
 {QCD Radiative Correction to Zero Recoil
Sum Rules for Heavy Flavor Transitions
     in the Small Velocity Limit.}\\[2cm]
\rm
 J.G.K\"{o}rner$^{1}$, K.Melnikov$^{2}$ and O.Yakovlev$^{3*}$\\
Johannes Gutenberg-Universit\"at\\
Institut f\"ur Physik, Staudingerweg 7\\
D-55099 Mainz, Germany\\
\normalsize
%\bf
%Abstract\\
%\rm
%\small
\end{center}
\begin{abstract}
  We consider the small velocity sum rules for heavy flavour
semileptonic transitions
that are used to  estimate the zero recoil values of
semileptonic heavy flavour form factors.
We analyze the complete O($\alpha _S$)
radiative correction to these sum rules.
The corrections are universal and influence all
"model-independent" bounds previously derived for
semileptonic form factors at zero recoil.\\[5cm]
\end{abstract}
\footnotesize
$^1$ Supported in part by the BMFT, Germany, under contract 06MZ730 \\
$^2\,$Supported by the Graduiertenkolleg Teilchenphysik, Universit\"at Mainz\\
$^3$ Supported by Deutsche Forschunggemeinschaft\\
$^{*)}$ on leave of absence from
Budker Institute for Nuclear Physics, Novosibirsk, Russia \\[2cm]

\normalsize
\end{titlepage}
\vspace{3cm}
{\bf 1.Introduction.}\\
An accurate determination of the Kobayashi-Maskawa matrix element
$V_{cb}$ is one of the most important
tasks of the heavy quark theory in the physics of heavy hadrons.
As is well known there are at least two possibilities for measuring
$V_{cb}$: i)
to measure inclusive semileptonic decays of heavy hadrons or ii) to
extrapolate  differential decay distributions of exclusive semileptonic
transition to the zero
recoil point that gives us
$\mid V_{cb}\mid f_A$. In the second method
precise theoretical predictions for the values of
form factor $f_A$ at zero recoil are extremely important
for extracting $V_{cb}$ from experiments.
\par
A general approach for obtaining such predictions
was suggested by Shifman et al. [1,2].
It is based on recent progress in the analysis of
inclusive semileptonic decays of
hadrons containing one heavy quark,
where the operator product expansion
(OPE)  method and HQET were applied [3-7].
The leading order result agrees with the free
b-quark decay picture.
Nonperturbative corrections start to appear
only at order
$(\frac{\Lambda_{QCD}}{M_Q})^2$ and
are determined by the matrix elements of only
a few local operators such as
the operators of the chromo-magnetic and the kinetic energy
 [3-7].
\par
The procedure of Shifman et al. [1,2] consists
in considering moments of spectral
distributions in the small velocity (SV) limit which allows one to obtain
corrections to known sum rules as well as new sum rules.
A very important result of this approach is that now one has an
 estimate of the
deviation of the value of the exclusive form factor $f_A$
from unity at
zero recoil.
\par
 The aim of the present note is to analyze  the $O(\alpha _S)$ corrections to
the SV sum rules.
Note that part of these corrections had
already been incorporated in the original derivation of the
sum rules.
We are referring to the vertex renormalization factors $
\eta _{A,V} $ which correspond to the finite renormalization of
the vertices
$ \langle c| \bar c \gamma _{\mu} b |b \rangle $ and $ \langle c | \bar c
\gamma _{\mu} \gamma _5 b |b \rangle $, respectively.
To the best of our knowledge these factors were first
introduced in [3]. However, we would like
to emphasize
that the vertex renormalization
 factors are not the whole story at the $O(\alpha _S)$
level and extra care is needed to derive correct sum rules with $O(\alpha _S)$
accuracy. The essential point here is that the diagrams with two partons in
the intermediate state are also involved.
\par
{\bf 2. Derivation of SV sum rules.}\\
 Let us begin with the forward scattering amplitude:
\begin{equation}
 T_{\mu \nu}(qv)=-i\int dx  e^{-iqx} <H_b\mid j_{\mu}(x)^*
j_{\nu}(0)\mid H_b>
\end{equation}
Here $j_{\nu}$ is the appropriate current and $v$ is
the velocity of
the heavy hadron $H_b$.
The function $T(qv)$ is an analytic function in the $(qv)-$
complex plane with the appropriate cuts. The structure of
the $T_{\mu \nu}$ cuts
is shown in Fig.1.
The cut $0\le qv \le M_{H_b}-M_{H_c}$ corresponds to
the decay of heavy hadron $H_b\to H_c+\nu l$
while the lower cut $qv<0$ and the upper cut $qv>M_{H_b}+
M_{H_c}$ represent other (crossed) physical processes.
Following Ref.[1] we shall argue latter on that, using duality concepts,
the contributions from the latter "crossed" cuts to the sum rules can
be neglected in as much as the theoretical and phenomenological contributions
on these cuts can be equated to one another.
The imaginary part of the forward scattering
amplitude on the "physical cut" $0\le qv \le M_{H_b}-M_{H_c} $
determines the hadronic tensor
$W_{\mu\nu}=-\frac{1}{\pi}ImT_{\mu\nu}$ which in turn determines
the inclusive
decay width of the hadron.
\par
Before discussing the calculation of the $
O(\alpha _s)$ corrections we would like to mention that
there exist two different approaches for deriving SV sum rules.
One was proposed by Chay, Georgi and Grinstein (CGG approach) [4] and
the other by Shifman et. al. (BSVU approach) [1].
Both approaches are based on the duality idea (global and local)
but use a
somewhat different language. We shall discuss them in turn.
\par
 The basis of  the CGG approach is  connected with
the possibility to perform an analytic continuation of the
forward scattering
amplitude to the whole complex $(qv)-$plane and to connect the
integral over the
physical cut with the integral over a "large-radius" circle in the complex
plane where the OPE is justified.
This can be thought of as a formal statement of the assumed
duality. A representative integration path $C_1$ is
shown in Fig.1 .\par
Integrating over the physical cut and equating  moments
of theoretical spectral functions
with their phenomenological counterparts  we obtain the sum rule:
\begin{equation}
\int _{phys.cut}^{}d \bar \epsilon ({\bar \epsilon})^n
 {W^{phen.}( \bar \epsilon )}
=
\int _{}^{}d \bar \epsilon ({\bar \epsilon})^n {W^{QCD}
( \bar \epsilon )},
\end{equation}
where $\bar \epsilon =qv-M_{H_b}-M_{H_c} $.
The integrand on the r.h.s of these sum rules includes
nonperturbative $\frac{1}{m_Q}$
power corrections as well as  radiative QCD corrections
which can be systematically incorporated by
using  standard OPE calculations.
\par
In the second (BSVU) approach [1] one writes down a
dispersion relation representation for the
forward scattering amplitude
\begin{equation}
T(\epsilon)=\frac {1}{\pi } \int_{phys.cut}
d \bar \epsilon \frac {ImT( \bar \epsilon
)}{ \epsilon - \bar \epsilon },
\end{equation}
and expands the integrand in powers of $ \frac {1}{\epsilon },
$ where, in
the rest frame of the initial hadron,  $ \epsilon = m_b - m_c-q_0 $.
Such an expansion is justified only for $\epsilon$-values
 $\epsilon >> \Lambda _{QCD} $  {\em and} $ \epsilon << 2m_c$.
However, in general there are excited states
with large invariant mass ($\approx  m_c, (m_b-m_c)$) whose
contributions are not suppressed (the perturbative analogue is surely
"hard" gluon emission).
This means that one cannot expand the integrand
of Eq.(3) in terms of powers of $\frac{1}{\epsilon}$.
%over the whole range of $\epsilon$.
It is then inevitable to split the region of integration
in the dispersion integral at some
scale $\mu >>\Lambda_{QCD}$ and $\mu <<2m_c$.
%The region of small
%$\mu$ corresponds to QCD nonperturbative region and other to perturbative
%one.
Rewriting Eq.(3) as
\begin{equation}
\int _{0}^{M_{H_b}-M_{H_c}}d \bar \epsilon \frac {W( \bar \epsilon )}
{ \epsilon - \bar \epsilon}=
\int _{0}^{\mu}d \bar \epsilon \frac {W( \bar \epsilon )}
{ \epsilon - \bar \epsilon}+
\int _{\mu}^{M_{H_b}-M_{H_c}}d \bar \epsilon \frac {W( \bar \epsilon )}
{ \epsilon - \bar \epsilon}.
\end{equation}
and assuming $\epsilon >> \mu $,
 we can expand the first term on the r.h.s. in terms of powers of
$\frac {1}{\epsilon}$ while
 the second term on the r.h.s. (which  reflects the contribution of
excited states) does not generally have a Laurant
expansion for $\epsilon << \mu$. However,
at this point one can invoke
local duality
%(which works well for enough large $\mu$)
to evaluate the second
term in the r.h.s.. In some sense this
piece of the theoretical spectral function has to incorporate
all radiative corrections coming from the hard region.
In other words we apply duality to equate the integral over the region
$[\mu, M_{H_b}-M_{H_c}]$ on the partonic side to the sum over excited states
(with energy $E>
\mu$). Thus we assume that duality is valid up to the scale $\mu<<m_c$.
Then doing an expansion in $\epsilon$ we obtain the result:
\begin{equation}
\int _{0}^{\mu}d \bar \epsilon ({\bar \epsilon})^n
 {W^{Phen.}( \bar \epsilon )}
=
\int _{0}^{\mu}d \bar \epsilon ({\bar \epsilon})^n {W^{QCD}
( \bar \epsilon )}.
\end{equation}
We want to emphasize that $\mu$ is the scale where
the excited states come into play. In this sense the$\mu$ scale
is quite similar to the energy of the continuum threshold in the usual
QCD/SVZ sum rules. It is well known that a particular choice for
the onset of
the
continuum contributions affects
the final result
of the QCD/SVZ sum rules analysis and typically results in
20-30 percent uncertainties.
This shows that some care is also needed to estimate the size of this
uncertainty in the analysis of the SV sum rules.\par
Remember that, when calculating QCD radiative correction to
the Wilson coefficients, we have to introduce yet another scale
which divides the region of integration into a perturbative
and a nonperturbative region. In order to distinguish these scales
we shall denote latter scale as $\mu_{OPE}$
whereas we denote the duality scale discussed above by $\mu_D$.
 From perturbative QCD we roughly expect $\mu _{OPE} \ge 300 MeV$
whereas we take $\mu _{D} \sim 1 GeV$ for the duality scale.
By definition the sum rules (2)-(5) cannot
depend on the scale $\mu_{OPE}$.
Also the SV sum rules do not depend on $\mu_{D}$ to leading
order in $\alpha_S$ and to any arbitrary
order in $\frac{1}{m_Q}$
 because the spectral functions originate entirely from the resonance region.
The leading order spectral function (leading in $\alpha_S$)
consists of a $\delta (\epsilon)$ function
and $\delta $-function derivatives.
Note though that, in the next to leading
order in $\alpha_S$, the spectral function starts to depend
on $\mu_{D}$ explicitly.
\par
Thus it is impossible to derive realistic "model-independent" bounds for the
form factors: the choice of "switching on" the
$\mu_{D}-$ dependence is intrinsically model dependent.
 It is worth mentioning that the same problem appears in the
CGG approach and is, technically, connected with the possibility to choose
different integration contours
in the $(qv)$-plane, each contour $C_2$ being defined by
the point where it leaves
the physical cut - this is
the aforementioned $\mu_{D}$ ambiguity (see Fig.1)
in a different guise.
\par
It is important to realize that the dependence on
$\mu_D$ appears only at next to leading order
in $\alpha_S$,
and not at leading order
as in the usual QCD/SVZ sum rules.
To estimate the size of this ambiguity and thereby to understand the
accuracy of the SV sum rules we have to compute QCD radiative corrections
to the SV
sum rules at arbitrary values of the scale $\mu_{D}$.
\par
{\bf 3. Results for structure functions.}\\
 In this section we present the result of calculating QCD radiative
corrections to those spectral functions that are
needed for the zero velocity sum rules.
The hadronic tensor $W_{\mu\nu}$ can be expanded in terms of 14
structure functions (see [4,6,8]):
\begin{eqnarray}
W_{\mu\nu}=-g_{\mu\nu}W_1+v_{\mu}v_{\nu}W_2-i\ep v^{\al}q^{\be}W_3
+q_{\mu}q_{\nu}W_4+(q_{\mu}v_{\nu}+q_{\nu}v_{\mu})W_5\\ \nonumber
-qs[-g_{\mu\nu}W_6+v_{\mu}v_{\nu}W_7-i\ep v^{\al}q^{\be}W_8
+q_{\mu}q_{\nu}W_9+(q_{\mu}v_{\nu}+q_{\nu}v_{\mu})W_{10}]\\ \nonumber
+(s_{\mu}v_{\nu}+s_{\nu}v_{\mu})W_{11}
+(s_{\mu}q_{\nu}+s_{\nu}q_{\mu})W_{12}+
i\ep v^{\al}s^{\be}W_{13}+i\ep q^{\al}s^{\be}W_{14}
\end{eqnarray}
where $v$ is the velocity and $s$ is the spin of the initial hadron.\par
As a next step one defines diagonal helicity structure functions
$W_L$ (longitudinal), $W_{T_{L,R}}$ (transverse left,right),
and $W_0$ (time-component or scalar)
by contracting the hadronic tensor $W_{\mu\nu}$
with the appropriate polarization vectors
${n_{\nu}}^{\lambda}{n_{\nu}}^{*\lambda}$ ($\lambda =0,+,-,t$).
(see Ref.[8] for details ).\par
At zero recoil one finds:
\begin{eqnarray}
W_L&=&W_1 \nonumber \\
W_0=-W_1+W_2&+&{(qv)}^2W_4+2(qv)W_5 \\
W_{T_{L,R}}=W_1 &\pm & \hat n \vec s W_{TS},\\ \nonumber
\end{eqnarray}
where
\begin{equation}
W_{TS}=(W_{13} + qvW_{14})
\end{equation}
and $\hat n $ defines the quantization axis of the off-shell $W$-boson.
\par
Next we calculate the zero recoil $O(\alpha_S)$ contributions
 to the  helicity
structure functions $W_0$, $W_L$ and $W_{TS}$.
The generic graphs which contribute to the $O(\alpha _s)$
correction are shown in Fig.2.
There are in principal several possibilities to cut the
graphs in Fig.2. The difference is the number of partons in the
intermediate state. The
cuts with one parton in the intermediate state reproduce the
vertex renormalization $\eta _{A,V}$
factors which were discussed earlier. The two-parton
cuts give rise to the absorptive contributions
in the range $0 \le qv \le m_b-m_c$.
We mention that the infrared singularites cancel in
the sum of the two-parton intermediate state contributions
in the soft gluon limit $qv \to m_b-m_c $.
This is in accord with the observation that the virtual
corrections have no infrared singularity at the zero recoil point [3,10].
\par
We first consider the correlator of two axial vector  currents
$j_{\mu}=\bar c\gamma_{\mu}\gamma_{5}b$ and obtain the following
zero recoil contributions:
\begin{equation}
{W_L}^{AA}(t)= \eta^2_A W_L^{Born}(t)+ \frac{\alpha_SC_F}{2\pi}
\frac{ (5t^2 + 10tx + 3x^2 + 2x + 3)(t + 2x)t}
{6(t + x)^3m_b},
\end{equation}
\begin{equation}
{W_0}^{AA}(t)=\eta^2 _A W_0^{Born}(t)+ \frac{\alpha_SC_F}{2\pi}
\frac{ (3t^2 + 6tx + x^2 + 2x + 1)(t + 2x)t}
{2(t + x)^3m_b},
\end{equation}
\begin{equation}
{W_{TS}}^{AA}(t)= \eta^2_A
W_{TS}^{Born}(t)+\frac{\alpha_SC_F}{2\pi}
\frac{ (t^2 + 2tx + x^2 - 2x - 3)(t + 2x)t}
{6(t + x)^3m_b},
\end{equation}
where $t=\frac{m_b-m_c-qv}{mb}$ and $x=\frac{m_c}{m_b}$.
The vertex renormalization factor
in Eq.(12) reads [3]:
\begin{eqnarray}
\eta _A = 1-\frac {\alpha _s}{\pi}(\frac {1+x}{1-x}log(x)+\frac {8}{3})
\end{eqnarray}
and, for the corresponding vector current case,
\begin{eqnarray}
\eta _V = 1-\frac {\alpha _s}{\pi}(\frac {1+x}{1-x}log(x)+2).
\end{eqnarray}
The $O(\alpha_S^0)$
Born contributions (which are state-dependent) include both the zeroth
order result and the effects of non-perturbative corrections and can be found
elsewhere [1,2,6-9].
Concerning the vector current case the structure functions (10)-(12)
 are the same except
for a sign change in the
fourth term of the first brackets ($2x$-term) in all formulas (10-12).\\

{\bf 4. QCD Radiative corrections to SV sum rules.} \\
 Next let us discuss the $O(\alpha _s)$ corrections to the SV sum rules.
As concerns sum rule applications the three helicity structure functions
${W_{L}}^{AA}, {W_{TS}}^{AA}$ and ${W_{0}}^{VV}$ are the most important. These
are the structure functions that have nonvanishing contributions at zero
recoil from the quasi-elastic contributions $B\to D,D^{*}$ and $
\Lambda_b \to \Lambda_c$ and $\Omega_b\to \Omega_c,\Omega_c^{*}$.
We want to discuss how these sum rules are modified when $O(\alpha _s)$
radiative corrections are taken into account.
\par
For the  "time-component" helicity structure function $W_0^{VV}$ we
obtain
\begin{equation}
(\int_{0}^{\mu_{OPE}}+\int_{\mu_{OPE}}^{\mu_D}) {W^{VV}_0(t)} dt =
\eta_{V}+ \frac {\alpha _sC_F}{2\pi }
(J_1(\mu_{D})-J_1(\mu_{OPE}))+n/pert(\mu _{OPE}),
\end{equation}
where $J_1$ is defined in the Appendix. The last term
in Eq.(15) is a symbolic notation
for the non-perturbative parameterization
of the contribution from the soft region: $[0,\mu_{OPE}]$.
Since we are mainly interested in how previously derived sum rules
change when the new $O(\alpha_s)$ corrections are included
the last term in Eq.(15) will not be written out explicitly in the
following since these contributions have been investigated in previous
papers (see Refs[1,2,8,9]).
On the other hand, $ \int _{0}^{\mu_{D}} {W _{0}}^{VV} dx $ is connected
with the sum rule for the vector form factors,
e.g. in the $ \Lambda _b $ case
(for form factor definitions see e.g. [8]):
\begin{equation}
|\sum \limits_{i=1}^{3} f_{V}^i |^{2} \le \int _{0}^{\mu_{D}} {W _{V}}^{(0)} dx
\end{equation}
Let us just for illustrative purposes assume
$\mu _{OPE} \to 0$ and $\mu_{D}=m_b-m_c$ that corresponds to integration
over whole physical cut. We then obtain:
%\begin{equation}
%\int _{0}^{m_b-m_c} {W^{VV} _0} dt = \eta_{V} + \frac {\alpha _sC_F}{8\pi }
%\cdot ( log(x)(5x^2+2x-1)-(x^4+x^3-x-1)).
%\end{equation}
%With the above simplifying assumptions
%the sum rule with $O(\alpha _s)$ accuracy reads:
\begin{equation}
|\Sigma f_{V} |^{2} \le  \eta _{V}+ \frac {\alpha _sC_F}{8\pi }
\cdot ( log(x)(5x^2+2x-1)-(x^4+x^3-x-1)).
\end{equation}
But integrating over resonance region $\mu_{D}<<m_c$
and assuming $\mu _{OPE} \to 0$ we have
\begin{equation}
|\sum \limits_{i=1}^{3} f_{V}^i
 |^{2} \le  \eta _{V}+ \frac {\alpha _sC_F}{4\pi }
(x-1)^2\frac{\mu_d^2}{m_c^2} +n/pert(0).
\end{equation}
Now let us consider the SV sum rules more carefully and estimate their
dependence on the duality scale $\mu_{D}$ and $\mu_{OPE}$.
The basic function $J_1(\mu)$ is shown in the Fig.3.
Let us emphasize that
$J_1$ is small
in the infrared region and hence
our result for perturbative correction is not sensitive to $\mu_{OPE}$.
%It is interesting here to estimate dependence of noperturbative operators
%on $\mu_{OPE}$. Using simple fact that r.h.s of Eq(15) does not depend
%on it we have
%\begin{equation}
%n/pert(\mu_{OPE})-n/pert(0)=\frac {\alpha _sC_F}{2\pi }
%J_1(\mu_{OPE})\to \frac {\alpha _sC_F}{12\pi }
%\frac{\mu^2_{OPE}}{m^2_c}(x^2-2x-3).
%\end{equation}
%Considering an expression for $n/pert$-term [1,2,7,8] we conclude that
%only kinetic energy depends on $\mu_{OPE}$:
%\begin{equation}
%\mu^2_{\pi}(\mu)-\mu^2_{\pi}(0)=\frac{\alpha_sC_F}{\pi}\mu^2.
%\end{equation}
%This quadratic dependence coincides with one obtained in Ref.[1].
Thus we may set $\mu_{OPE}=0$ and forget about the $\mu_{OPE}$-dependence of
operators with higher dimensions
and use the known expressions for the $\eta_{V,A}$ factors [3].
\par
In order to make reliable estimates we have to decide on the value of
$\mu_{D}$. In principle using the CGG approach
we may choose any
$\mu_{D}$ in the interval $[\mu_{OPE},m_b-m_c]$
based on various assumptions about the
applicability of duality. Conventionally
one takes $\mu_D\simeq 1-3$ GeV in QCD sum rule applications.
%May we put ($\mu_{OPE}=\mu_{duality}$?
We see from Fig.3 that the result for the
QCD radiative corrections to leading
operator does have a substantial dependence on $\mu_D$  in the region 1-3 GeV.
The size of the effect varies from 0.5
$\%$
at $\mu_D=1$GeV to 3.
$\%$
at $\mu_D=3$GeV,
here we have used $\alpha_s=0.3$, $m_b=4.8$Gev , $m_c=1.5$GeV for definiteness.
\par
A similar situation occurs for all the other sum rules
considered in the literature before. For instance, taking
the external current to be axial and projecting on the longitudinal helicity
function for $B \to D^*$, the radiative corrections
considered here change the original inequality
$\mid f_A \mid^2\le \mid \eta_A\mid^2+n/pert$  derived in Refs. [1,2]
to
\begin{equation}
| f_{A} |^{2} \le  \eta _{A}+ \frac {\alpha _sC_F}{2\pi }
\cdot (J_2(\mu_{D})-J_2(\mu_{OPE}))+n/pert(\mu_{OPE})
\end{equation}
with $J_2$ from the Appendix.
In the case $\mu _{OPE} \to 0$ and $\mu_{D}<<m_c$ we obtain
\begin{equation}
|f_{A} |^{2} \le  \eta _{A}+ \frac {\alpha _sC_F}{4\pi }
(x^2+\frac{2}{3}x+1)\frac{\mu_d^2}{m_c^2}+ n/pert(0).
\end{equation}
This result changes the prediction for the bound $f_{A}<0.94$
obtained in
Ref.[2]. The bound is pushed upward by 0.5$\%$ and 2.$\%$ for $\mu_D=1$
and $3$ GeV, respectively.
\par
Note that the authors of Refs. [1,2] have
used one single scale $\mu=\mu_D=
\mu_{OPE}$ implicitly including considered effects
to nonperturbative operators.
But if this is the case one has
to estimate and include the $\mu$-dependence of
the nonperturbative matrix elements.
This is important at $O(\alpha _s)$ accuracy since
the $\mu$-dependent part of these operators
turns out to be of the order $\alpha_S(\frac{\mu}{m_c})^2\approx \alpha_S$
at $\mu\approx m_c$.
A simple way to avoid this problem is to take $\mu <<m_c$ as it has
been done
in Ref.[1,2]. However, strictly speaking, this choice for the value of $\mu$
is not quite harmless as
one applies duality concept
up to an extremely  small scale $\mu<<m_c$ where local duality can break
down (see for example Ref.[11]).
Moreover, if we use the commonly accepted
numerical values for $\Lambda _{QCD}$
and $m_c$ then the reliability of the strong inequality is doubtful:
\begin{equation}
\Lambda _{QCD} << \mu << m_c
\end{equation}
\par
On the other hand a numerical analysis of our formulas
shows that the typical size of radiative
corrections to the small velocity sum rules is of the order
of 1$\%$ at $\mu=m_c$.
We can regard this as an estimate of the theoretical uncertainty within
the SV sum rule calculations.
Phenomenologically this results in a $1-2\%$ increase of the relevant bounds
on the zero recoil form factors.
\par
Applying SV sum rules techniques one can also obtain inequalities for the
matrix elements of non-perturbative operators.
For example for the
$\Lambda _b $-case discussed in Ref.[8] the following inequality was obtained:
\begin{equation}
{\mu _s}^2+\frac {{\mu _{\pi}}^2}{3} \le 0
\end{equation}
Here ${\mu _{\pi}}^2$ stands for the expectation value of the heavy
quark kinetic energy operator while ${\mu _{s}}^2$ parameterizes the
$\frac {1}{{m_b}^2}$ correction for the axial vector current matrix element
between $\Lambda _b$ states.\par
Using our previous results the inequality (21) gets changed to
\begin{equation}
{\mu _s}^2+\frac {{\mu _{\pi}}^2}{3} - \frac{\alpha_SC_F}{2\pi}m_b^2
(J_3(\mu_{D})-J_3(\mu_{OPE}))
\le 0
\end{equation}
when the $O(\alpha_S)$ radiative corrections are included. The function
$J_3$ is defined in the Appendix.
Again at $\mu _{OPE} \to 0$ and $\mu_{D}<<m_c$ we obtain
\begin{equation}
{\mu _s}^2+\frac {{\mu _{\pi}}^2}{3} +
\frac {\alpha _sC_F}{4\pi }
(1+\frac{2}{3}x-\frac{1}{3}x^2)m_b^2\frac{\mu_D^2}{m_c^2}
\le 0 .
\end{equation}
All terms in the last equation can be of the same order of
 magnitude in principle ($\mu^2_{\pi}\approx 0.5 GeV$,
 last term in Eq.(24) is about $0.8(\frac{\mu_D}{m_c})^2$)
and there are no a priori reason
to discriminate between
them. But then the values for $ {\mu _s}^2 $ and $ {\mu _{\pi}}^2
$ are not connected directly  what is the major advantage
when the radiative corrections are neglected.\par
{\bf 5. Conclusion }\\
To conclude, we have analysed the $O(\alpha _s) $ radiative corrections
to zero
recoil sum rules for  semileptonic heavy hadrons form factors.
Our results are  universal and shift all previously
derived model-independent bounds
on zero recoil form factors by 1-2 percents
with obvious consequences for the extraction
of $ V _{cb} $ from experiment.

{\bf 6. Acknowledgments}  When this work was completed
the extended version
of Ref.[1] appeared where similar questions were discussed.
The results obtained in Ref.[1] are in agreement with our results.
We would like to thank A.Vainshtein
for clarifying conversations on the subject.

\begin{center}{\bf 7. Appendix}
\end{center}
 In this appendix we collect our results for the $O(\alpha_S)$
QCD corrections to the leading order term
in the SV sum rules.
These corrections arise from the two parton intermediate states.\\
The QCD radiative correction to the zeroth moment
of the ${W_0}^{VV}$
helicity structure
function
 ($J_1(\mu)=\int_{0}^{\mu}W_{0}^{VV}(t)dt$)
in the
vector current  correlator is
\begin{eqnarray}
J_1(\mu)=
\frac{1}{4(y+x)^2}(-2(yx+log(\frac{y+x}{x})
(y+x)^2)(5x^2+2x-1)\\ \nonumber
-y^2(17x^2+2x-1)
-3y^4-12y^3x)
\end{eqnarray}
with $y=\mu /m_b$ and $x=\frac{m_c}{m_b}$.
For the ${W_L}^{AA}$ helicity structure function the
correction to zeroth moment is given by:
\begin{eqnarray}
J_2(\mu)=
\frac{1}{12(y+x)^2}(-2(yx+log(\frac{y+x}{x})
(y+x)^2)(7x^2-2x-3)\\ \nonumber
-y^2(27x^2-2x-3)
-5y^4-20y^3x).
\end{eqnarray}
Finally for the transverse structure function ${W_{TS}}^{AA}$ one has
\begin{eqnarray}
J_3(\mu)=\frac{1}{12(y+x)^2}(-2(yx+log(\frac{y+x}{x})
(y+x)^2)(x^2+2x+3)\\ \nonumber
-y^2(5x^2+2x+3)
-y^4-4y^3x).
\end{eqnarray}

\begin{center}{\bf 8. References.}
\end{center}
1. I.Bigi, M.Shifman, N.Uraltsev, A.Vainstein TPI-MINN/12-T.\\
2. M.Shifman, N.Uraltsev, A.Vainstein TPI-MINN/13-T.\\
3. M. Voloshin and M. Shifman, Yad. Fiz. 47 (1988) 801\\
\mbox{\rule{5mm}{0mm}}[Sov.J.Nucl.Phys.47(1988) 511].\\
4. J.Chay, H.Georgi, B.Grinstein, Phys. Lett. B247 (1990) 399.\\
5. I.Bigi, M.Shifman, N.Uraltsev, A.Vainstein, Phys. Rev. Lett. 71,(1993) 496
\\
6. B.Blok, L.Koyrakh, M.Shifman, A.Vainstein, Phys. Rev. D49 (1994) 3356. \\
7. A.Manohar, M.Wise , Phys. Rev. D49 (1994) 1310. \\
8. J.G. K\"{o}rner, D.Pirjol, Phys. Lett. B334 (1994) 399.\\
9. J.G. K\"{o}rner, K. Melnikov, O. Yakovlev
   preprint MZ-TH/94-24, September 94. \\
10. M. Neubert, Nucl. Phys. B 371, 149 1992.\\
11. B.Blok,R. Dikemann and M. Shifman Preprint TPI-MINN-94/23-T.\\
\mbox{\rule{5mm}{0mm}}[hep-ph/9405246].\\
%\end{document}

\newpage
\begin{picture}(0,0)%
%%%\special{psfile=disk$userinst:[yakovlev]fig1.pstex}%
\end{picture}%
\setlength{\unitlength}{0.012500in}%
\begin{picture}(275,336)(46,485)
\end{picture}
\newpage
\begin{picture}(0,0)%
%%%\special{psfile=disk$userinst:[yakovlev]fig2.pstex}%
\end{picture}%
\setlength{\unitlength}{0.012500in}%
\begin{picture}(290,377)(5,428)
\end{picture}
\newpage
\begin{figure}
\epsfxsize=15cm
%%%\centerline{\epsffile{figg1.psfix}}
\end{figure}
\newpage
%\documentstyle[12pt]{article}
%\begin{document}
\begin{center}{\bf Figure captions}
\end{center}

{\bf Fig.1} The structure of the forward scattering cuts in the $qv$
 complex plane. $C_1$ and $C_2$ are representative integration paths. \\

{\bf Fig.2} The generic graphs which contribute to the $O(\alpha_S)$
correction. \\

{\bf Fig.3} Dependence of $J_1(\mu)$ (upper), $J_2(\mu)$
 and $-J_3(\mu)$ (lower on r.h.s.)
on $\mu$ at $m_b=4.8GeV$, $m_c=1.5GeV$.\\

\end{document}